\documentclass[twocolumn,amsmath,nofootinbib]{revtex4} 
\usepackage{url}
\usepackage{amssymb}
\usepackage{amsfonts}
\usepackage{amsbsy}
\usepackage{graphicx}
\usepackage{hyperref}
\usepackage{array} 
\usepackage{multirow}

\newcolumntype{x}[1]{%
>{\centering\hspace{0pt}}p{#1}}%
\providecommand{\openone}{\leavevmode\hbox{\small1\kern-3.8pt\normalsize1}}

\def\ie{{\frenchspacing\it i.e.}}
\def\eg{{\frenchspacing\it e.g.}}
\def\etc{{\frenchspacing\it etc.}}

\def\spose#1{\hbox to 0pt{#1\hss}}
\def\simlt{\mathrel{\spose{\lower 3pt\hbox{$\mathchar"218$}}
   \raise 2.0pt\hbox{$\mathchar"13C$}}}
\def\simgt{\mathrel{\spose{\lower 3pt\hbox{$\mathchar"218$}}
     \raise 2.0pt\hbox{$\mathchar"13E$}}}
 \def\simpropto{\mathrel{\spose{\lower 3pt\hbox{$\mathchar"218$}}
     \raise 2.0pt\hbox{$\propto$}}}

\def\beq#1{\begin{equation}\label{#1}}
\def\eeq{\end{equation}}
\def\beqa#1{\begin{eqnarray}\label{#1}}
\def\eeqa{\end{eqnarray}}
\def\eq#1{equation~(\ref{#1})}	
\def\Eq#1{Equation~(\ref{#1})}
\def\eqn#1{~(\ref{#1})}

\def\fig#1{Figure~\ref{#1}}
\def\Fig#1{Figure~\ref{#1}}

\def\Sec#1{Section~\ref{#1}}
\def\Sec#1{Section~\ref{#1}}

\def\ed{\end{document}}

\def\lambdavec{{\boldsymbol\lambda}}
\def\muvec{{\boldsymbol\mu}}

\def\bfzero{{\bf 0}}

\def\I{{\bf I}}
\def\A{{\bf A}}
\def\B{{\bf B}}
\def\C{{\bf C}}
\def\E{{\bf E}}
\def\F{{\bf F}}
\def\H{{\bf H}}

\def\P{{\bf P}}

\def\SS{{\bf S}}
\def\s{{\bf s}}
\def\U{{\bf U}}
\def\V{{\bf V}}

\def\Hsubj{H_{\rm s}}
\def\Hobj{H_{\rm o}}
\def\Henv{H_{\rm e}}
\def\Hso{H_{\rm so}}
\def\Hoe{H_{\rm oe}}
\def\Hse{H_{\rm se}}

\def\tr{\hbox{tr}\,}

\def\tensormult{\otimes}
\def\expec#1{\langle#1\rangle}

\def\ket#1{|#1\rangle}

\def\bra#1{\langle #1|}

\def\ketbra#1#2{\left|#1\right\rangle\left\langle#2 \right|}

 \def\ns{\mskip-5mu}
 
\def\rn{}
\def\nn#1 #2{#2. #1}				
\def\nnn#1 #2 #3{#2. #3. #1}			
\def\nnnn#1 #2 #3 #4{#2. #3. #4 #1}		
\def\nnnnn#1 #2 #3 #4 #5{#2. #3. #4 #5. #1}	
\def\dualand{ and\hbox{ }}				
\def\multiand{, and\hbox{ }}				
\def\rf#1;#2;#3;#4;#5 {{\frenchspacing\par\rn#1, #3 {\bf #4}, #5 (#2). \par}}
\def\rg#1;#2;#3;#4;#5;#6 {{\frenchspacing\par\rn#1, #3 {\bf #4}, #5 (#2). \par}}
\def\rfbook#1;#2;#3;#4;#5 {{\frenchspacing\par\rn#1, {\it #3} (#5, #4, #2).\par}}
\def\rfprep#1;#2;#3 {{\par\frenchspacing\rn#1, #3 (#2).\par}}
\def\rfproc#1;#2;#3;#4;#5;#6 {{\frenchspacing\par\rn#1 #2, in {\it #3}, ed. #4 (#5: #6)\par}}
\def\rfprocp#1;#2;#3;#4;#5;#6;#7 {{\frenchspacing\par\rn#1 #2, in {\it #3}, ed. #4 (#5: #6), p#7\par}}

\begin{document}
\pdfoptionalwaysusepdfpagebox=5
\abovedisplayshortskip=4pt
\belowdisplayshortskip=4pt
\abovedisplayskip=3.0pt
\belowdisplayskip=3.0pt



\title{Sharpening the Second Law of Thermodynamics with the Quantum Bayes' Theorem}

\author{Hrant Gharibyan}
\author{Max Tegmark}

\address{Dept.~of Physics \& MIT Kavli Institute, Massachusetts Institute of Technology, Cambridge, MA 02139}

\date{\today}
\date{To appear in Physical Review E; submitted October 3 2013, accepted August 20 2014}

\vspace{10mm}

\begin{abstract}
We prove a generalization of the classic Groenewold-Lindblad entropy inequality, combining 
decoherence and the quantum Bayes theorem into a simple unified picture
where decoherence increases entropy while observation decreases it. 
This provides a rigorous quantum-mechanical version of the second law of thermodynamics, governing how the
entropy of a system (the entropy of its density matrix, partial-traced over the environment and conditioned on what is known) evolves under general decoherence and observation. 
The powerful tool of spectral majorization enables both simple alternative proofs of the classic Lindblad and Holevo inequalities without using strong subadditivity, and also novel inequalities for decoherence and observation that hold not only for von Neumann entropy, but also for arbitrary concave entropies.  
\end{abstract} 
 
\maketitle

\section{Introduction}
\vskip-3mm

More than a century after its formulation, the second law of thermodynamics 
remains at the forefront of physics research, with continuing progress  on 
generalizing its applications 
and
clarifying its foundations. 
For example, it is being extended to non-equilibrium statistical mechanics \cite{Bunin11},
quantum heat engines \cite{SkrzypczykShortPopescu13},
biological self-replication \cite{England12}
and cosmological inflation \cite{tripartite}.
As to the quantum-mechanical version of the second-law, 
Seth Lloyd showed that can be derived from imperfectly known quantum evolution \cite{LloydThesis},
and one of us showed how it can be generalized to observed open systems \cite{brain,tripartite}.
The goal of the present paper is to complete this generalization by providing the required mathematical proofs.

\begin{figure}[pbt]
\centerline{\includegraphics[width=80mm]{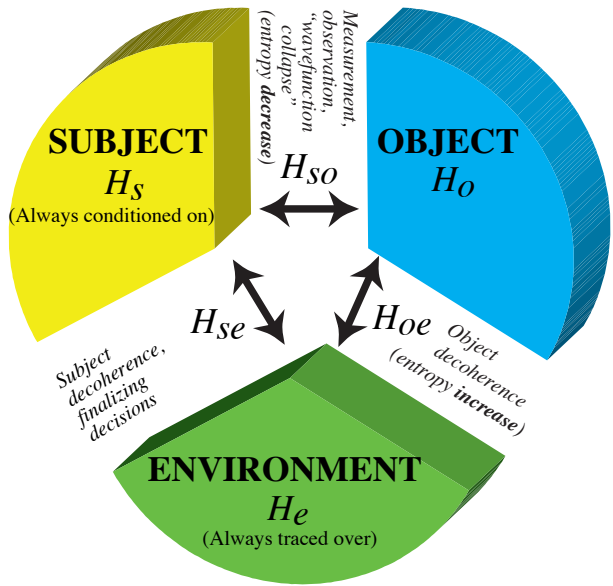}}
\vskip-3mm
\caption{
``EOS-decomposition'' of the world. 
The subsystem Hamiltonians 
$\Hsubj$, $\Hobj$, $\Henv$ and 
the interaction Hamiltonians
$\Hso$, $\Hoe$, $\Hse$ can
cause qualitatively different effects,
providing a unified picture including both observation and decoherence.
}
\vskip-16mm
\label{TrinityFig}
\end{figure}

As emphasized by Von Neumann \cite{VonNeumann32} and Feynman \cite{Feynman72}, 
state of a quantum system is completely described by a density matrix $\rho$, which encodes everything 
we need to know to make the best possible predictions for its future behavior.
However, if you are interested in using physics to make predictions about your own future, knowing $\rho$ for 
the entire universe (or multiverse) is neither sufficient nor necessary \cite{tripartite}.
\vspace{0mm}
\begin{enumerate}
\itemsep-1mm
\item{\bf Is not sufficient:} You also need to know what branch of the global wavefunction you are in. In particular, you need to take into account what you know about your location, both in 3D space and in Hilbert space.
\item{\bf Is not necessary:} You only need to know the quantum state ``nearby'', both in 3D space and in Hilbert space. 
\end{enumerate}
\vspace{-2mm}
To predict what you will observe in a quantum physics lab, you need to take into account both which of the many existing physics labs you happen to be in and also which quantum state preparation has been performed. On the other hand, you do not need to take into account the current state of the Andromeda Galaxy or branches of the wavefunction that have permanently decohered from your own. 

To quantify this, you can always decompose the total system (the entire cosmos) into three subsystems as illustrated in \fig{TrinityFig}: the degrees of freedom corresponding to your subjective perceptions (the subject), the degrees of freedom being studied (the object), and everything else (the environment). Computing the correct density matrix for your object of interest therefore involves two steps \cite{tripartite}:
\vspace{-2mm}
\begin{enumerate}
\itemsep-1mm
\item{\bf Condition} on what you know (on all subject degrees of freedom).
\item {\bf Marginalize}  over what you don't care about (partial-trace over all environment degrees of freedom).
\end{enumerate}
\vspace{-2mm}
The first step, including quantum state preparation and observation, can be thought of as the quantum generalization of Bayes' Theorem \cite{tripartite}.
The second step produces decoherence, helping explain the emergence of a classical world \cite{Zeh70,ZehBook,Zurek09,SchlosshauerBook}.
As we will see below, decoherence always increases entropy whereas observation on average decreases it.
Although the latter was proved by Shannon for the case of classical physics, the corresponding quantum theorem has hitherto eluded proof except in spacial cases. We will sharpen the second law as follows:
\vspace{-2mm}
\begin{enumerate}
\itemsep-1mm
\item{\bf Observation:} When an object is probed by the subject, its entropy on average decreases.
\item {\bf Decoherence:}  When an object is probed by the environment, its entropy increases.
\end{enumerate}
\vspace{-1mm}
Below we will define ``probed'' and provide rigorous mathematical proofs of these entropy inequalities.

\begin{table*}
\caption{ Summary of the two basic quantum processes discussed in the text.
    \label{BigPicture}
}
\def\arraystretch{1.9}
\begin{tabular}{|p{3.7cm}|c|c|}
\hline
{\bf For observer to predict future, global $\rho$... }
   &    ...is not sufficient 	        	&      ...is not necessary    \\
\hline
{\bf Mathematical Operation}&  Condition (on  subject degrees of freedom)	&  Marginalize (over environment degrees of freedom) \\
\hline
{\bf Interaction}   &  Object-Subject  &  Object-Environment \\
\hline
{\bf Process}  &  Observation	&  Decoherence  \\
\hline
{\bf Dynamics}  &  $\rho_{ij}\mapsto \rho_{ij}^{(k)} = \rho_{ij}{S_{ik} S_{jk}^*\over p_k},\quad p_k\equiv \sum_i\rho_{ii}|S_{ij}|^2$ &  $\rho_{ij}\mapsto\rho_{ij}E_{ij}$	\\
\hline
{\bf Entropy Inequality} & Decrease: $\sum_k p_k S\left(\rho^{(k)}\right) \le S(\rho)$  & Increase: $S(\rho) \le S\left(\rho\circ E \right)$  \\
\hline
\end{tabular}
\vskip-3mm
\end{table*}

\def\zupstate{
\setlength{\extrarowheight}{2pt}
\left(
\begin{tabular}{cc}
$1$	&$0$\\
$0$	&$0$
\end{tabular}
\right)
}

\def\xupstate{
\setlength{\extrarowheight}{2pt}
\left(
\begin{tabular}{cc}
${1\over 2}$	&${1\over 2}$\\
${1\over 2}$	&${1\over 2}$
\end{tabular}
\right)
}

\def\mixedstate{
\setlength{\extrarowheight}{2pt}
\left(
\begin{tabular}{cc}
${1\over 2}$	&$0$\\
$0$		&${1\over 2}$
\end{tabular}
\right)
}
%

\section{Effects of decoherence and observation}
\vskip-4mm
Let us begin by briefly reviewing the unitary cosmology formalism of \cite{brain,tripartite} necessary for our proofs.
We define  {\it probing}
as a nontrivial interaction between the object and some other system that leaves the object unchanged in some basis,  \ie, such that the unitary dynamics $\U$ merely changes the 
system state in a way that depends on the object state $\ket{o_i}$.We will justify and generalize this definition in \Sec{InequalitySec}.\\
\\
%
{\bf Object-Environment:} If the object is probed by the environment the unitary dynamics of object-environment system is given by
\beq{InteractionAssumptionEq}
\U\ket{o_i} \ket{e_*}=\ket{o_i}\ket{\epsilon_i}
\eeq
where $\ket{e_*}$ and $\ket{\epsilon_i}$ denote the initial and final states of the environment for the object state $\ket{o_i}$. The density matrix describing the object is the object-environment density matrix partial-traced over the environment. \Eq{InteractionAssumptionEq} implies that the object density matrix $\rho$ evolves as \cite{Zeh70,tripartite} 

\beq{DecoEq}
\rho \mapsto \rho\circ\E,
\eeq
where the matrix $\E$ is defined by $E_{ij}\equiv \langle\epsilon_j| \epsilon_i\rangle$ and the symbol 
$\circ$ denotes what mathematicians know as the {\it Schur product}.
Schur multiplying two matrices simply corresponds to multiplying their corresponding components, \ie, 
$(\rho\circ\E)_{ij}=\rho_{ij} E_{ij}$.\\
\\
{\bf Object-Subject:} If the object is instead probed by an observer the unitary dynamics of the object-subject system is given by 
\beq{InteractionAssumptionEq}
\U\ket{o_i}\ket{s_*}=\ket{o_i} \ket{\sigma_i}
\eeq
where $\ket{s_*}$ and $\ket{\sigma_i}$ denote the initial and final states of the subject for the object state $\ket{o_i}$. Let $\ket{s_k}$ denote the basis states that the subject can perceive, which are robust to decoherence (discussed in \cite{tripartite} and \cite{Tegmark2014}) and will correspond to ``pointer states" \cite{ZurekHabibPaz93} for the case of human observer. Since the subject will rapidly decohere, the observer will with probability $p_k$  find  that her state is $\ket{s_k}$ and that the object density matrix is $\rho^{(k)}$, where \cite{tripartite}
\beq{ObsEq}
\rho^{(k)} = {\rho\circ (\s^k \s^{k\dagger})\over p_k},\quad
p_k\equiv \sum_i\rho_{ii}|s^k_i|^2.
\eeq
Here the vector $\s^k$ is the $k^{\it th}$ column of 
the matrix $S_{ik}\equiv \langle s_k| \sigma_i\rangle$, \ie, 
$s^k_i\equiv S_{ik}$ and $\rho_{ij}^{(k)} = \rho_{ij}{S_{ik} S_{jk}^*/p_k}$.
\Eq{ObsEq} can be though of as the quantum-mechanical version of Bayes' Theorem \cite{tripartite}, showing how the observer's state of knowledge about a system gets updated by conditioning on new observed information.

These effects of decoherence and observation are intimately related. 
Since the states  $|s_k\rangle$ form a basis, let us define
$F_{ij}= \langle\sigma_j| \sigma_i\rangle=
\sum_k \langle\sigma_j   |s_k\rangle  \langle s_k    |\sigma_i\rangle=
\sum_k S_{jk}^* S_{ik}$, i.e.  
\beq{FSeq}
\F=\SS\SS^\dagger.
\eeq
\Eq{ObsEq} and (\ref{FSeq}) imply that 
\beq{DecoObsRelationEq}
\rho\circ\F = \sum_k p_k\rho^{(k)}.
\eeq
This means that decoherence (eq.~\ref{DecoEq}) can be interpreted as an observation that we do not know the outcome of: 
after being probed by its environment (setting $\F=\E$), the object is in one of the states $\rho^{(k)}$ (with probability $p_k$), 
and we simply do not know which. 

The most revealing 
probing is when $\SS=\E=\I$, the identity matrix: then the system gains complete information about the object, so 
decoherence makes $\rho$ diagonal (so-called von Neumann reduction \cite{VonNeumann32}) and observation produces pure states $\rho^{(k)}=|\tilde{o}_k\rangle\langle \tilde{o}_k|$, for some orthonormal basis $\ket{\tilde{o}_k}$. 
Another interesting special case is when all elements $S_{ij}$ are zero or unity.
Then the decoherence equation\eqn{DecoEq} reduces to what is known as L\"uders projection \cite{Lueders1950}
\beq{LuedersEq}
\rho\circ\E = \sum_k\P_k \rho\P_k,
\eeq
and the observation equation\eqn{ObsEq} gives
$\rho^{(k)}\propto\P_k\rho\P_k$,
where $\P_k\equiv \sum_i S_{ik}|o_i\rangle\langle o_i|$
are orthogonal projection operators (satisfying $\P_i\P_j=\delta_{ij}\P_i$, $\sum\P_i=\I$).
The least revealing probing is the trivial case when 
$\rho^{(k)}$ and $\rho\circ\E$ equal $\rho$ up to a unitary transformation, 
so that the subject or environment learns nothing about the object. We define ``probing'' to exclude this trivial case, which occurs for example when $\SS$ is of the form
$S_{jk}=e^{i(\theta_j+\phi_k)}$ or, for the decoherence case, when $\rho$ is diagonal.
\vskip-3mm

\section{Entropy inequalities (S-theorems)}
\label{InequalitySec}
\vskip-3mm

We will now prove the main result of this paper: that observation on average decreases entropy, while decoherence increases entropy. Specifically, we will prove the theorem 
\beq{EntropyTheorem}
\sum_k p_k S\left(\rho^{(k)}\right) \le S(\rho) \le S\left(\sum_k p_k \rho^{(k)}\right),
\eeq
relating the expected entropy after observation (left), the initial entropy (middle) and the entropy after decoherence (right). Both ``$\le$'' become ``$<$" when the probing is nontrivial. We will refer to these entropy inequalities as two S-theorems. Our proof below holds for the general types of observation from \eq{ObsEq} and decoherence from \eq{DecoEq}, and for a very general definition of entropy: for any quantify of the form
\beq{EntropyDefEq}
S(\rho)\equiv\tr h(\rho)
\eeq
where $h$ is a concave function on the unit interval ($h''(x)<0$ for $0\le x\le 1$). This includes the Shannon/von Neumann entropy ($h(x)=-x\ln x$), the linear entropy ($h(x)=1-x^2$),
the rescaled exponentiated Renyii entropy ($H(x)=\pm x^\alpha$) and the log-determinant ($h(x)=\ln x$). 

\begin{figure}[pbt]
\centerline{\includegraphics[width=90mm]{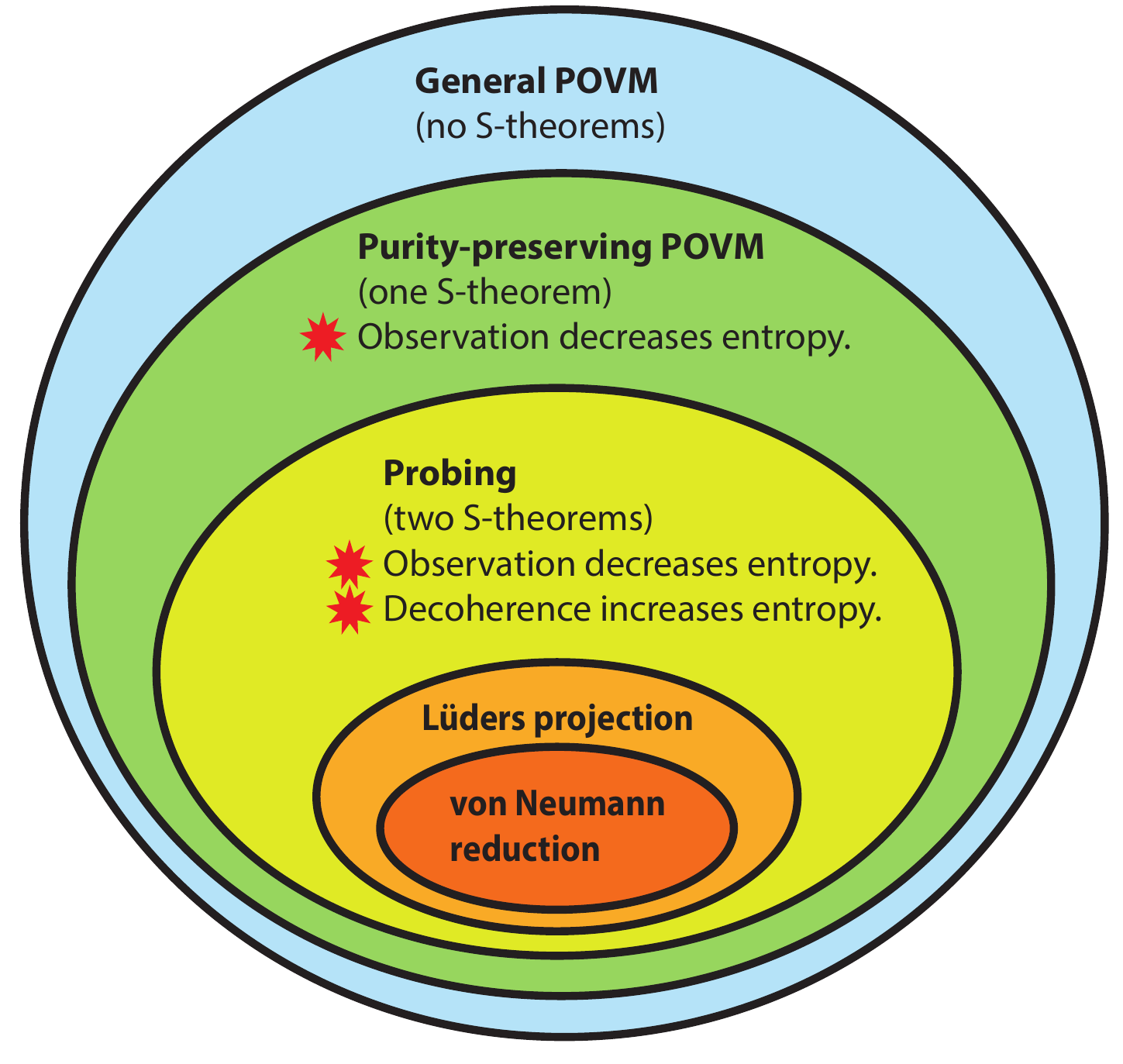}}
\vskip-3mm
\caption{Our two entropy theorems generalize previous results from complete measurements (von Neumann reduction) and projective measurements (L\"uders projection) to probing, the most general interaction that can be interpreted as an object measurement. 
The observation S-theorem further generalizes to purity-preserving POVMs, whereas general POVMs violate both S-theorems.
}
\label{POVM}
\end{figure}

In Appendix~\ref{GeneralMeasurement}, we study interactions that are more general than probing, obtaining the results illustrated in \Fig{POVM}.
By providing counterexamples that violate both inequalities in \eq{EntropyTheorem}, we prove that for a general 
Positive Operator Valued Measure (POVM) \cite{PreskillNotes}, neither of the two S-theorems holds true. 
We also show that our proof of the left (observation) inequality of \eq{EntropyTheorem} can be generalized from probing to what we term a purity-preserving POVM, 
``PPPOVM", the special type of POVM that maps any pure state into pure states.
We find that for PPPOVMs, observation on average decreases entropy, but decoherence does {\it not} always increase entropy.
The interactions we defined as probing are simply the subset of PPPOVMs that leave the object unchanged in at least one basis.
These are the most general interactions in \Fig{POVM} that can be interpreted as measurements.
To see this, consider that simply replacing the object state by some fixed and {\it a priori} known pure state (as in counterexample 2 in Appendix~\ref{GeneralMeasurement}) is a PPPOVM, and it would be ridiculous to view this as a measurement of the object.

\subsection{Majorization and entropy}
\vskip-3mm

Our proof uses numerous inequalities involving the notion of  {\it majorization} \cite{MarshallBook}, which we will now briefly review. One writes 
\beq{MajorizationNotationEq}
\lambdavec\succ\muvec
\eeq
and says that the vector $\lambdavec$ with components $\lambda_1$, ...,$\lambda_n$ {\it majorizes} the vector $\muvec$ with components 
$\mu_1$, ...,$\mu_n$ if they have the same sum and 
\beq{MajorizationDefEq}
\sum_{i=1}^j \lambda_i \ge \sum_{i=1}^j \mu_i\quad\hbox{for }j=1,\dots,n,
\eeq
\ie, if the partial sums of the latter never beat the former:
$\lambda_1\ge\mu_1$, 
$\lambda_1+\lambda_2\ge\mu_1+\mu_2$, {\etc} 
It is not difficult to show (see, \eg,  \cite{MarshallBook} or Appendix A of \cite{tripartite}) that if $\lambdavec\succ\muvec$,
then
\beq{ConcavityTheorem}
\sum_i h(\lambda_i) \le \sum_i h(\mu_i)
\eeq
for any concave function $h$.
This means that if the vectors are probability distributions (so that $\lambda_i\ge 0$, $\sum_i\lambda_i=1$)
with entropies defined as
\beq{lambdaEntropyEq}
S(\lambdavec)=\sum_i h(\lambda_i),
\eeq
then
the majorization $\lambdavec\succ\muvec$ implies the entropy inequality 
$S(\lambdavec)\le S(\muvec)$.
Letting $\lambdavec(\rho)$ denote the eigenvalues of the density matrix $\rho$ sorted in decreasing order, and using \eq{EntropyDefEq}, we thus have the powerful result 
\beq{MajorizationEntropyRelalation}
\lambdavec(\rho_1)\succ\lambdavec(\rho_2)    \implies    S(\rho_1)\le S(\rho_2),
\eeq
so if two density matrices $\rho_1$ and $\rho_2$ satisfy
$\lambdavec(\rho_1)\succ\lambdavec(\rho_2)$, then they satisfy the entropy inequality $S(\rho_1)\le S(\rho_2)$.

\vskip-80mm

\subsection{The proof} \label{TheProof}
\vskip-3mm

The right part of the entropy theorem\eqn{EntropyTheorem} (that decoherence increases entropy) was proven in \cite{tripartite}.
By \eq{DecoObsRelationEq}, it is equivalent to 
\beq{DecoTheoremEq}
S(\rho\circ\E)\ge S(\rho),
\eeq
which follows from\eqn{MajorizationEntropyRelalation} and the majorization  
\beq{MajorizationEq}
\lambdavec(\rho\circ\E)\prec\lambdavec(\rho),
\eeq
which is Corollary J.2.a in \cite{MarshallBook} (their equation 7), which in turn follows from a 1985 theorem by Bapat and Sunder. 
To prove the other half of  the entropy theorem\eqn{EntropyTheorem} (that observation decreases entropy), we will need the following theorem, which is proven in Appendix~\ref{ProofAppendix}:
For any Hermitean matrix $\rho$ and any complete orthogonal set of projection operators $\P_i$ (satisfying 
 $\sum_i\P_i=\I$, $\P_i\P_j = \delta_{ij}\P_i$, $\P_i^\dagger=\P_i$), 
\beq{DoubleInequalityTheorem}
 \sum_i\lambdavec \left(\P_i\rho\P_i\right)
 \succ
 \lambdavec(\rho) 
 \succ
 \lambdavec\left(\sum_i\P_i\rho\P_i\right).
\eeq
Applying \eqn{MajorizationEntropyRelalation} to the 
left part gives
\beqa{JensenEq}
S(\rho)&\ge&S\left(\sum_k\lambdavec \left(\P_k\rho\P_k\right)\right)
=S\left(\sum_k p_k\lambdavec\left({\P_k\rho\P_k\over p_k}\right)\right)=\nonumber\\
&=&\sum_i h\left(\sum_k p_k\lambdavec_i\left({\P_k\rho\P_k\over p_k}\right)\right)\ge\nonumber\\
&\ge& \sum_{ik} p_kh\left(\lambdavec_i\left({\P_k\rho\P_k\over p_k}\right)\right)\nonumber
= \sum_k p_k S\left({\P_k\rho\P_k\over p_k}\right),
\eeqa
where we used Jensen's inequality in the penultimate step.
This, \eqn{DoubleInequalityTheorem} and\eqn{MajorizationEntropyRelalation} thus
imply that for any density matrix $\rho$,
\beq{EntropyLemma}
\sum_i p_i S\left({\P_i\rho\P_i\over p_i}\right)\le S(\rho) \le S\left(\sum_i{\P_i\rho\P_i}\right),
\eeq
where $p_i\equiv\tr \P_i\rho\P_i$.
The right half of this double inequality is an alternative proof of \eq{DecoTheoremEq} for the
special case where decoherence is a L\"uders projection as in \eq{LuedersEq}.
Using instead the left half of \eqn{EntropyLemma}, we obtain the following proof of the remaining (left) part of our 
entropy theorem\eqn{EntropyTheorem}:
\beqa{ObsEntropyProofEq}
\expec{S}&\equiv&\sum_k p_k S(\rho^{(k)})=
\sum_k p_k S(\rho^{(k)} \tensormult |s_k\rangle\langle s_k|)=\nonumber\\
&=&\sum_k p_k S\left({\P_k \U\rho^*\U\P_k\over p_k}\right)
\le S(\U\rho^*\U) =\nonumber\\
&=&S(\rho^*)=S(\rho\tensormult |s_*\rangle\langle s_*|)=S(\rho).
\eeqa
Here $\rho^*\equiv\rho\tensormult |s_*\rangle\langle s_*|$ is the initial state of the combined object-observer system,
which the observation process evolves into $\U\rho^*\U$, which in turn decoheres into 
$\sum_k\P_k \U\rho^*\U\P_k=\sum_k p_k\rho^{(k)} \tensormult |s_k\rangle\langle s_k|)$ as derived in \cite{tripartite},
where 
$\P_k\equiv \I \tensormult |s_k\rangle\langle s_k|$ are projection operators acting on the combined object-observer system.
The first and last equal signs in \eq{ObsEntropyProofEq} hinge on the fact that tensor multiplying a density matrix by a pure state leaves its entropy unchanged, merely augmenting its spectrum by a number of vanishing eigenvalues.
The inequality step uses\eqn{EntropyLemma}, and the subsequent step uses the fact that unitary evolution leaves entropy unchanged.



\vskip-3mm
\section{Conclusions}
\vskip-3mm

We have proved the entropy inequality\eqn{EntropyTheorem}, which states that
decoherence increases entropy whereas observation on average decreases it. 
Its left half is a direct generalization of the Groenewold-Lindblad inequality \cite{Lindblad1972}, which corresponds to the special case of the projective (L\"uders) form of measurement and the special case of von Neumann entropy; our results also hold Renyii entropy, linear entropy and indeed any entropy defined by a concave function.
Both of these entropy inequalities hold for interactions {\it probing} the object, defined as the most general interactions leaving the object unchanged in some basis. Of the various classes of interactions we considered, probing constitutes the most general one that can be interpreted as a measurement of the object. We showed that our observation inequality, but not our decoherence inequality, holds also for more general interactions that are purity-reserving POVMs, generalizing Ozawa's result for quasi-complete measurements beyond von Neumann entropy \cite{Ozawa86}. 
None of our entropy inequalities hold for arbitrary POVMs. 

To prove inequalities (\ref{EntropyTheorem}), we used the link between unitary quantum mechanics and spectral majorization. This link was first noticed by Uhlmann \cite{Uhlmann71} and later proved extremely helpful in the study of pure state transformations and entanglement \cite{Nielsen99}. We showed that majorization techniques can also be used to provide simple alternative proofs of entropy inequalities for observation and decoherence.
Also it complements Holevo's inequality \cite{Holevo1973}
$\sum_k p_k S(\rho_k)\le S(\rho)$, where $\rho=\sum_k p_k\rho_k$, 
which follows immediately from \eq{SumMajorizationTheorem} and Jensen's inequality.
It also generalizes Shannon's classical version thereof, which states that observation on average reduces the entropy by an amount equal to the mutual information. 
The quantum entropy reduction cannot always be that large (which would give negative entropy after observing a qubit with $\SS=\I$ and two bits of mutual information), but we have proved that it is never negative.

Our results complete the formalism of \cite{brain,tripartite} for handling
observed open systems. We have seen that quantum statistical mechanics still works flawlessly as long as we avoid 
sloppy talk
of {\it the} density matrix and {\it the} entropy: we each have our own personal density matrix encoding everything we know about our object of interest, and we have simple formulas (Table~\ref{BigPicture}) for how it changes under observation and decoherence, decreasing and increasing entropy.

{\bf Acknowledgments:}
The authors wish to thank Seth Lloyd, Robert Penna, Benjamin Schumacher, Harold Shapiro and Wojciech Zurek for helpful discussions.
This work was supported by NSF grant AST-1105835 and the MIT UROP program.

\vskip-3mm

\appendix
\section{Proof of \Eq{DoubleInequalityTheorem}}
\label{ProofAppendix}
\vskip-3mm

Let us first review three useful facts that we will use in our proof.
As proven in \cite{Bourin11},
any Hermitean matrix $\H$ written in block form can be decomposed as 
\beq{DecompositionTheorem}
 \H= 
 \left[ 
  \begin{array}{ c c }
	\A & \C \\
	\C^\dagger &\B
  \end{array} \right]
  = 
  \U \left[ 
  \begin{array}{ c c }
	\A & \bfzero \\
	\bfzero & \bfzero
  \end{array} \right] \U^\dagger
  +
  \V \left[ 
  \begin{array}{ c c }
	\bfzero & \bfzero \\
	\bfzero & \B
  \end{array} \right] \V^\dagger
\eeq
for some unitary matrices $\U$ and $\V$.
Second, for any two Hermitean matrices $\A$ and $\B$, 
\beq{SumMajorizationTheorem}
\lambdavec(\A+\B) \prec \lambdavec(\A) + \lambdavec(\B).
\eeq
This theorem was suggested and proved by Fan in 1949 and the proof is provided in \cite{Fuzhen2011} as Theorem 10.21.
Finally, because the spectrum of a matrix is invariant under unitary transformations, we have
\beq{UnitarityEq}
\lambdavec(\U\H\U^\dagger)=\lambdavec(\H)
\eeq
for any Hermitean matrix $\H$ and any unitary matrix $\U$.

Combining these three facts, we obtain
\beqa{ProofInequality1}
\lambdavec\left(
\begin{tabular}{cc}
$\A$&$\C$\\
$\C^\dagger$&$\B$
\end{tabular}
\right)
&=&
\lambdavec\left[
\U\left(
\begin{tabular}{cc}
$\A$&$\bfzero$\\
$\bfzero$&$\bfzero$
\end{tabular}
\right)\U^\dagger
+
\V\left(
\begin{tabular}{cc}
$\bfzero$&$\bfzero$\\
$\bfzero$&$\B$
\end{tabular}
\right)\V^\dagger
\right]
\prec\nonumber\\
&\prec&
\lambdavec\left[
\U\left(
\begin{tabular}{cc}
$\A$&$\bfzero$\\
$\bfzero$&$\bfzero$
\end{tabular}
\right)\U^\dagger\right]
+
\lambdavec\left[\V\left(
\begin{tabular}{cc}
$\bfzero$&$\bfzero$\\
$\bfzero$&$\B$
\end{tabular}
\right)\V^\dagger
\right]=\nonumber\\
&=& 
\lambdavec\left(
\begin{tabular}{cc}
$\A$&$\bfzero$\\
$\bfzero$&$\bfzero$
\end{tabular}
\right)
+
\lambdavec\left(
\begin{tabular}{cc}
$\bfzero$&$\bfzero$\\
$\bfzero$&$\B$
\end{tabular}
\right),
\eeqa
where the three logical steps use equations~(\ref{DecompositionTheorem}), (\ref{SumMajorizationTheorem}) and (\ref{UnitarityEq}), respectively.

Now consider a complete orthogonal set of Hermitean projection operators 
$\P_i$, $i=1,...,n$, satisfying the standard relations $\sum_{i=1}^n\P_i=\I$ and $\P_i\P_j=\delta_{ij}\P_i$. 
Since $\P_i^2 = \P_i$, all eigenvalues are either $0$ or $1$.
Since all matrices $\P_i$ commute, there is basis where they are all diagonal, and where
each matrix vanishes except for a block of ones somewhere along the diagonal.
In this basis, $\P_i\H\P_i$ is simply $\H$ with all elements set to zero except for a corresponding square block.
For example, for $n=2$ we can write
\beq{PdefEq}
\P_1\ns\left(\ns
\begin{tabular}{cc}
$\A$&$\C$\\
$\C^\dagger$&$\B$
\end{tabular}
\ns\right)
\ns\P_1
=
\left(\ns
\begin{tabular}{cc}
$\A$&$\bfzero$\\
$\bfzero$&$\bfzero$
\end{tabular}
\ns\right),
\quad
\P_2\ns\left(\ns
\begin{tabular}{cc}
$\A$&$\C$\\
$\C^\dagger$&$\B$
\end{tabular}
\ns\right)
\ns\P_2
=
\left(\ns
\begin{tabular}{cc}
$\bfzero$&$\bfzero$\\
$\bfzero$&$\B$
\end{tabular}
\ns\right).
\eeq
This means that we can rewrite the inequality (\ref{ProofInequality1})
as
\beq{ProofInequality2}
\lambdavec(\H)\prec\sum_{i=1}^2 \lambdavec(\P_i\H\P_i)
\eeq
for any Hermitean matrix $\H$.
The sum of any of our two projection operators is a new projection operator, 
so by iterating \eq{ProofInequality2}, we can trivially generalize it to the case of arbitrary $n$:
\beq{ProofInequality3}
\lambdavec(\H)\prec\sum_{i=1}^n \lambdavec(\P_i\H\P_i).
\eeq
This concludes the proof of the left part of the double inequality (\ref{DoubleInequalityTheorem}).
The right part follows directly from the \eq{MajorizationEq} 
when making the choice 
\beq{EchoiceEq}
\E_{ij}\equiv\sum_{k=1}^n (\P_k)_{ii} (\P_k)_{jj}
\eeq
in the basis where all projectors are diagonal.
In other words, $\E$ is chosen to be the matrix with ones in all square blocks picked out by the projectors and zeroes everywhere else.

\vskip-6mm

\section{POVMs and Measurements}
\label{GeneralMeasurement}
\vskip-3mm

In this appendix, we derive the extent to which our entropy inequalities can be extended to interactions more general than probing.
A Positive Operator Valued Measure (POVM) \cite{PreskillNotes} is a projective measurement on the larger system that has the object as a subsystem. The additional quantum system is typically referred to as an ancilla (A). Specifically, a POVM is a mapping
\beq{RhoMap}
\rho \rightarrow \{ p_k, \rho^{(k)}\}, 
\eeq 
where the resulting object state is $\rho^{(k)}$ with probability of outcome $p_k$. \\

\subsection{General POVM}

 $\{ p_k, \rho^{(k)}\}$ are given by
\beq{G-POVM}
\rho^{(k)} = {\chi_k(\rho) \over \tr \chi_k(\rho)}, \quad p_k =  \tr \chi_k(\rho),
\eeq 
where 
\beq{G-POVM2}
\chi_k(\rho) = \tr_A \left( \P^{(k)}_{OA} \left[ \U \rho \tensormult \rho_A\U^\dagger \right] \right).
\eeq
Here $A$ denotes the ancilla system in some initial state $\rho_A$ and $\{ \P^{(k)}_{OA}\}$ is orthonormal set of projectors acting on the object-ancilla system. Without loss of generality, we can take the ancilla state to be pure ($\rho_A=\ket{a^*}\bra{a^*}$), because for a mixed ancilla state
$\rho_A=\sum_i w_i \ket{a_i}\bra{a_i}$, the purified state $\ket{a^*}\equiv \sum_i w_i^{1/2} \ket{a_i}\tensormult\ket{a_i}$ defines the same POVM.
 
For a general POVM, neither of the two inequalities in the \Eq{EntropyTheorem} holds true, so we have no S-theorems for general POVMs. 
We will now prove this by providing POVM counterexamples that violate both the  
left (observation) and right (decoherence) inequalities in \Eq{EntropyTheorem}. 
\begin{itemize}
\item \label{Example1}
{\it Counterexample 1:} The object and ancilla are single qubits in a state $\ket{0}_O \ket{0}_A$ and the POVM is defined by $\U=\I$ and $\P^{(k)}_{AO} = \ketbra{\Psi_k}{\Psi_k}$, where 
\beqa{BellStates}
\ket{\Psi_1} &=& \frac{\ket{00}+\ket{11}}{\sqrt{2}}, \quad 
\ket{\Psi_2} = \frac{\ket{00}-\ket{11}}{\sqrt{2}} \\ 
\nonumber \ket{\Psi_3}&=& \frac{\ket{01}+\ket{10}}{\sqrt{2}}, \quad
\ket{\Psi_4} = \frac{\ket{01}-\ket{10}}{\sqrt{2}}. \\ 
\eeqa
This gives
\beqa{BellExample}
\rho^{(1)} &=& \rho^{(2)} = \rho^{(3)} = \rho^{(4)} =\mixedstate,\\
p_1 &=& p_2 = \frac{1}{2}, \quad \quad\quad p_3 = p_4 = 0,
\eeqa
so that the initial von Neumann entropy $S(\rho) =0$ increases to a final entropy of 1 bit, violating the left (observation) part of the \Eq{EntropyTheorem}. 

\item \label{Example2}
{\it Counterexample 2:} The object and ancilla are single qubits in initial states
\beqa{}
&\rho = \mixedstate \quad \text{and} \quad
&\rho_A = \zupstate,
\eeqa
and $\U$ is the unitary two qubit operation that exchanges the states of these qubits (a swap gate)
\beq{}
 \U \rho \tensormult \ket{0}\bra{0} \U^\dagger = \ket{0}\bra{0}  \tensormult  \rho.
\eeq
The projections $\P^{(k)}_{OA}$ are
\beqa{}
\P^{(1)}_{OA} &=& \I \otimes \ketbra{0}{0},\\ 
\P^{(2)}_{OA} &=& \I \otimes \ketbra{1}{1},
\eeqa
giving
\beqa{BellExample}
\rho^{(1)} &=& \rho^{(2)} = \zupstate, \\ \nonumber
p_1 &=& p_2 =  \frac{1}{2},
\eeqa
so that the initial entropy $S(\rho)=1$ bit drops to zero, violating the right (decoherence) side of the inequality (\ref{EntropyTheorem}). 
\end{itemize}

\subsection{Purity-preserving POVM (PPPOVM)}

 We define a PPPOVM as a POVM that keeps a pure state pure, i.e., if $\rho$ is a pure state, then $\rho^{(k)}$ is pure for all $k$. 
 A POVM is purity preserving if and only if
\beq{AncillaProject}
 \P^{(k)}_{OA}  = \I \otimes \ketbra{a_k}{a_k},
\eeq  
where $\ket{a_k}$ is an orthonormal basis of the ancilla system. To show this we, need to prove both the ``if" and ``only if" parts.
\begin{itemize}
\item
{\bf Part 1 (``if''):} For a pure object state $\rho = \ketbra{\psi}{\psi}$,  
the pure total state 
\beq{}
\ket{\Psi}^{OA}\equiv\U \ket{\psi} \ket{a^*} 
\eeq
can always be decomposed as
\beq{}
\ket{\Psi}^{OA} = \sum_k \lambda_k \ket{\psi_k} \ket{a_k},
\eeq  
where $\ket{\psi_k}$ are a normalized object states and  $\lambda_k$ are complex numbers. 
 If $ \P^{(k)}_{OA} = \I \otimes \ketbra{a_k}{a_k}$, then this 
decomposition gives
\beq{}
\chi_k(\rho) = |\lambda_k|^2 \ketbra{\psi_k}{\psi_k}.
\eeq
Since the state $\rho_k = \ketbra{\psi_k}{\psi_k}$ is pure for all $k$, we have shown that 
if $ \P^{(k)}_{OA}  = \I \otimes \ketbra{a_k}{a_k}$, then the POVM is purity preserving. 
\item
{\bf Part 2 (``only if''):} If the POVM is purity preserving, then we can write $\rho = \ketbra{\psi}{\psi}$ and $\rho_k = \ketbra{\psi_k}{\psi_k}$,  so we have
\beqa{}
\chi_k(\rho) &=& \tr_A \left( \P^{(k)}_{OA} \ket{\Psi}^{OA} \bra{\Psi}^{OA} \right) \\  \nonumber
&=& p_k \ketbra{\psi_k}{\psi_k}.
\eeqa
 This means that after normalization, the state $\P^{(k)}_{OA} \ket{\Psi}^{OA}$ must be pure and separable for all $k$:
\beq{}
\P^{(k)}_{OA} \ket{\Psi}^{OA} = \lambda_k \ket{\psi_k} \ket{a_k},
\eeq
where $|\lambda_k|^2 = p_k$ and $\ket{a_k}$ is some state of the ancilla system. We can now express $\P^{(k)}_{OA} $ as
\beq{}
\P^{(k)}_{OA}  = \I \otimes \ketbra{a_k}{a_k},
\eeq
and because these terms form an orthonormal set of projectors, we conclude that $\ket{a_k}$ form an orthonormal basis for ancilla. We have thus shown that if the POVM is purity preserving, then $ \P^{(k)}_{OA}  = \I \otimes \ketbra{a_k}{a_k}$ for some orthonormal basis $\ket{a_k}$.
\end{itemize} 

In \cite{Ozawa86},  Ozawa referred to PPPOVMs as ``quasi-complete''.
For the special case of von Neumann entropy for 1-dimensional position measurements, he
showed that the left inequality in \Eq{EntropyTheorem} is true if and only if a measurement is quasi-complete. As we will now show, our proof of the left part of \Eq{EntropyTheorem} is readily generalized to PPPOVMs for our much more general definition of entropy, 
and without use of strong subadditivity. Here the ancilla plays the role of the subject and projections on the object-ancilla system is given by \Eq{AncillaProject}, which are equivalent to the projectors $\P_k = \I \otimes \ketbra{s_k}{s_k}$ used in equations (\ref{DoubleInequalityTheorem}) and (\ref{ObsEntropyProofEq}). Because the key equations (\ref{DoubleInequalityTheorem}) and (\ref{ObsEntropyProofEq}) hold for arbitrary $\U$, the proof for the left (observation) part of \Eq{EntropyTheorem} can be immediately generalized to PPPOVMs. In contrary, the right part of \Eq{EntropyTheorem} does not hold for PPPOVM's, which can be seen using the counterexample 2 from the previous section. So only one of the S-theorems holds
for PPPOVMs: observation always decreases average entropy, but decoherence does not always increase it. \\

\subsection{Probing} 

Probing is a special case of a PPPOVM, where the role of the ancilla is played by either the subject or the environment, and the unitary dynamics $\U$ of the object-ancilla system leaves the object unchanged in some basis, merely changing the ancilla state in a way that depends on the object state $\ket{o_i}$:
\beq{}
\U \ket{o_i} \ket{a^*} = \ket{o_i} \ket{a_i}.
\eeq
In this paper, we have shown that both inequalities in \Eq{EntropyTheorem} hold true for probing, so that observation always decreases average entropy and decoherence always increases it.

\clearpage

\bibliographystyle{h-physrev}
\bibliography{main}

\end{document}